\documentclass[twocolumn,pra,showpacs]{revtex4}

\usepackage{amsmath,amssymb}
\usepackage{graphics}
\usepackage{epsfig}

\begin{document}

\title{Three-body systems with attractive $1/r$ potentials} 
\author{J. P. D'Incao, S. C. Cheng, H. Suno and B. D. Esry}
\affiliation{Department of Physics, Kansas State University,
Manhattan, Kansas 66506, USA} 

\begin{abstract}
We have used the hyperspherical adiabatic representation to describe
the system of three identical bosons in an spin stretched state
interacting by an attractive $1/r$ potential. A proposal has been made
how such a system might be realized experimentally in cold trapped
atoms using extremely off-resonant laser fields [Phys. Rev. Lett. 84,
5687 (2000)]. We have obtained effective potentials, channel functions,
and nonadiabatic couplings for this gravity-like interaction, allowing
us to calculate the ground state energy with accuracy that
substantially improves upon previous results. We have similarly
calculated the energies for the first four $0^{+}$ excited
states. These results show that the simple adiabatic hyperspherical
approximation offers an accurate description for such a system.
\end{abstract}

\maketitle 

\section{introduction}

Recently, a scheme for inducing gravitation-like interatomic
potentials has been proposed \cite{DODell2000}, opening up the
possibility to create self-bound Bose-Einstein condensates (BECs)
\cite{DODell2001,DODell2002,Choi,Hu,Giovanazzi}. In such a scheme, the
gravitation-like interatomic potential can be  
achieved by irradiating the atoms with intense, extremely off-resonant  
electromagnetic fields. The usual strong anisotropy due to
dipole-dipole interactions can, in fact, be averaged out \cite{Thiru}
by the proper combination of laser beams, leaving a $-u/r$ potential,
where $u$ is the strength of the potential (the analogue of $GMm$,
where $G$ is the Newton's constant and $M$ and $m$ are the masses) and
$r$ is the interparticle distance. The strength $u$ of the potential
can be adjusted by changing the laser intensity \cite{DODell2000}.  

In ultracold atomic gases, two interesting regimes for self-bound BECs
have been predicted, assuming that the short-range interatomic
interactions can be independently tuned by, say, applying a magnetic 
field near a Feshbach resonance \cite{Fesh}. 
In one regime, the
attractive $1/r$ interactions are balanced by the repulsive mean field
interation assuming a positive two-body scattering length and
negligible kinetic energy. In the other regime, the balancing factor
is kinetic energy, assuming negligible mean field interactions. In
both regimes, the resulting BEC is self-bound.
From a broader point of view, 
the induced gravitation-like interaction might make possible
experimental emulation of boson stars (a system of self-gravitating
bosons) in the regime where the kinetic energy balances the $-u/r$
potential \cite{Wang,IMMoroz}. Moreover, purely attractive $1/r$
potentials constitute an interesting contrast to the attractive {\em and}
repulsive Coulomb potentials atomic physics are used to.

The existence of a lower bound for the ground state energy in
many-body systems interacting through attractive $1/r$ potentials is
of fundamental importance in order to prove the existence of the
thermodynamical limit and the stability of normal matter
\cite{Fisher}. It was shown in Ref.~\cite{JLBasdevant} that for a
system of $N$ identical, spinless (or spin stretched)
bosons of mass $m$ interacting gravitationally, the lower and upper
bounds for the ground state energy are, respectively,
$-\frac{1}{16}N^2(N-1)\:G^2m^5/\hbar^2$ and
$-0.0542N(N-1)^2\:G^2m^5/\hbar^2$, where the upper bound was obtained
variationally. 
For small $N$, however, the discrepancy between the
lower and upper bounds becomes large. For $N=3$, using a more
refined trial function, they obtained $-0.95492\:G^2m^5/\hbar^2$
for the upper bound, representing a difference of about 
$15\%$ between the upper and lower ($-\frac{9}{8}\:G^2m^5/\hbar^2$),
bounds and a ground state energy equal to 
$E_{0}\cong -1.067\:G^2m^5/\hbar^2$. 

In this paper, we have used the adiabatic hyperspherical
representation for this systm to obtain effective three-body
potentials, the corresponding channel functions, and the nonadiabatic
couplings. Using these, we calculate 
the ground state and low-lying $0^{+}$ excited state energies
converged to seven digits. We have also used the adiabatic
hyperspherical representation to obtain lower and upper bounds that
differ by about $0.1\%$, indicating that a simple single-channel
description offers a quite accurate description of such systems.

\section{The adiabatic hyperspherical representation} 

We have solved the Schr\"odinger equation in hyperspherical
coordinates. After separation of the center-of-mass motion, the system
is described by the hyperradius $R$ which gives the 
overall size; three Euler angles $\alpha$, $\beta$ and
$\gamma$, specifying the orientation of the plane containing the three
particles relative to the space-fixed frame; and other two hyperangles
$\varphi$ and $\theta$, describing the internal relative motion
between the particles. We have defined $\varphi$ and $\theta$ as a
modification of Smith-Whitten coordinates
\cite{Esry-02,Esry-03,Smith}.     
The key to the adiabatic hyperspherical representation is that the
dynamics of the three-body system is reduced to collective motion
under the influence of one-dimensional effective potentials in $R$,
which is governed by a system of ordinary differential equations.

\begin{figure}
\includegraphics[width=2.in,angle=0,clip=true]{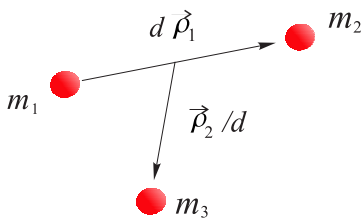}
\caption{The mass-scaled Jacobi coordinates for systems with three
  particles.}  
\label{JacobiCoords}
\end{figure}

The hyperspherical coordinates are introduced through the mass-scaled
Jacobi coordinates $\vec{\rho}_{1}$ and $\vec{\rho}_{2}$ (see
Fig. \ref{JacobiCoords}) defined as 
\begin{eqnarray}
&&\vec{\rho}_{1}=(\vec{r}_{2}-\vec{r}_{1})/d, \nonumber \\
&&\vec{\rho}_{2}=d\left(\vec{r}_{3}
-\frac{m_{1}\vec{r}_{1}+m_{2}\vec{r}_{2}}{m_{1}+m_{2}}\right).  
\end{eqnarray}
\noindent
In the above equations $\vec{r}_{i}$ is the position of the particle $i$
(of mass $m_{i}$) relative to a space-fixed frame. For three identical
particles of mass $m$, we define a three-body reduced mass as
$\mu=m/\sqrt{3}$ which gives $d=2^{1/2}/3^{1/4}$ \cite{Esry-02,Esry-03}. It is
important to note that the hyperradius,  
\begin{equation}
R^2=\rho_{1}^2+\rho_{2}^2, \hspace{0.25in}R\in[0,\infty), 
\end{equation}
\noindent 
is an invariant quantity, i.e., it does not depend on the particular
choice of the hyperangles or labels of the particles.  

The Schr\"odinger equation can be more conveniently written in terms
of the rescaled wave function $\psi=R^{5/2}\Psi$, as 
\begin{eqnarray} 
\left[-\frac{\hbar^2}{2\mu}\frac{\partial^2}{\partial R^2}
+H_{\rm ad}(R,\Omega)\right]\psi(R,\Omega)=E\psi(R,\Omega),\label{schr} 
\end{eqnarray} 
\noindent
where $E$ is the total energy  and $H_{\rm ad}$ is the adiabatic
Hamiltonian given by 
\begin{eqnarray}
H_{\rm ad}(R,\Omega)=\frac{\hbar^2}{2\mu R^2}
\left[{\Lambda^{2}(\Omega)+\frac{15}{4}}\right]+V(R,\varphi,\theta). 
\label{had}
\end{eqnarray} 
\noindent
The adiabatic Hamiltonian $H_{\rm ad}$ 
contains all hyperangular dependence, represented collectively by 
$\Omega\equiv\{\varphi,\theta,\alpha,\beta,\gamma\}$, and includes the  
hyperangular kinetic energy in the grand angular momentum operator
$\Lambda^{2}$ as well as all interparticle interactions $V$.  

In the adiabatic hyperspherical representation, the total wave 
function is expanded in terms of the channel functions
$\Phi_{\nu}(R;\Omega)$, 
\begin{equation}
\psi_{n}(R,\Omega)=\sum_{\nu}F_{n\nu}(R)\Phi_{\nu}(R;\Omega),
\label{chfun}
\end{equation}
\noindent 
where $F_{n\nu}(R)$ are the hyperradial wavefunctions, $n$ labels the
different energy eigenstates for a given $\nu$, and $\nu$ represents
all remaining quantum numbers necessary to specify each  
channel. The channel functions $\Phi_{\nu}(R;\Omega)$ form a complete 
set of orthonormal functions at each value of $R$ and are
eigenfunctions of the adiabatic Hamiltonian:   
\begin{equation}
H_{\rm ad}(R,\Omega)\Phi_{\nu}(R;\Omega)
=U_{\nu}(R)\Phi_{\nu}(R;\Omega).\label{poteq}
\end{equation}
\noindent
The eigenvalues $U_{\nu}(R)$ help define
effective three-body potentials 
for the hyperradial motion.  

Substituting Eq.~(\ref{chfun}) into the Schr\"odinger equation
(\ref{schr}) and projecting out $\Phi_{\nu'}$, 
we obtain the hyperradial Schr\"odinger equation  
\begin{widetext}
\begin{eqnarray}
\left[-\frac{\hbar^2}{2\mu}\frac{d^2}{dR^2}+U_{\nu}(R)\right]F_{\nu}(R) 
-\frac{\hbar^2}{2\mu}\sum_{\nu'}
\left[2P_{\nu\nu'}(R)\frac{d}{dR}+Q_{\nu\nu'}(R)\right]F_{\nu'}(R)
=EF_{\nu}(R),\label{radeq}
\end{eqnarray}
\end{widetext}
\noindent
which describes the motion of the three-body system under
the influence of the effective potentials
$U_{\nu}(R)-Q_{\nu\nu}(R)/2\mu$. The nonadiabatic coupling terms
$P_{\nu\nu'}(R)$ and $Q_{\nu\nu'}(R)$ drive inelastic collisions
three-body scattering processes and are defined as    
\begin{eqnarray} 
P_{\nu\nu'}(R) &=&
\Big\langle\hspace{-0.15cm}\Big\langle\Phi_{\nu}\Big|
\frac{d}{dR}\Big|\Phi_{\nu'}\Big\rangle\hspace{-0.15cm}\Big\rangle
\label{puv}
\end{eqnarray}
\noindent
and
\begin{eqnarray} 
Q_{\nu\nu'}(R) &=&
\Big\langle\hspace{-0.15cm}\Big\langle\Phi_{\nu}\Big|
\frac{d^2}{dR^2}\Big|\Phi_{\nu'}\Big\rangle\hspace{-0.15cm}\Big\rangle,
\label{quv}
\end{eqnarray} 
\noindent
where the double brackets denote integration over the angular
coordinates $\Omega$ only. As it stands, Eq.~(\ref{radeq}) is
exact. In practice, of course, the sum over channels must be
truncated. In fact, the accuracy of the solutions can be monitored
with successively larger truncations since the bound state
energies obtained at each stage are an upper bound by the variational
principle. 

In this paper, we explore the solutions of the system of differential  
equations (\ref{radeq}) for three particles with attractive $1/r$
interactions. We determine the effective potentials and couplings by
solving Eq.~(\ref{poteq}) for the $J^{\pi}=0^{+}$ symmetry, where $J$
is the total orbital angular momentum and $\pi$ is the total
parity. The low-lying bound state energies are then determined by
solving Eq.~(\ref{radeq}). 

In order to solve the adiabatic equation (\ref{poteq}), we have
expanded the channel functions $\Phi_{\nu}(R;\Omega)$ in terms of the 
Wigner $D$ functions \cite{Esry-02,Parker1987,RoseBook},
\begin{equation}
\Phi_{\nu}^{JM\pi}(R;\Omega)=\sum_{K}\phi_{K\nu}(R;\theta,\varphi)
D^J_{KM}(\alpha,\beta,\gamma), 
\end{equation}
where $K$ and $M$ are the projection of $\vec{J}$ onto the body-fixed
and space-fixed $z-$axes, respectively. After projecting out the $D$
functions, the resulting coupled system of partial differential
equations for $\phi_{K\nu}(R;\theta,\varphi)$ is solved (for each value
of $R$) by expanding $\phi_{K\nu}(R;\theta,\varphi)$ on a direct product
of fifth order basis splines \cite{Boor1978} in the hyperangles
$\theta$ and $\varphi$ \cite{Esry-02,Esry-03}. For $J^{\pi}=0^+$, of course,
the sum involves only one term, requiring the solution of a single
two-dimensional partial differential equation.  

The potential $V$ in Eq.~(\ref{had}) is given by a pairwise sum of
attractive $1/r$ potentials,   
\begin{equation}
V(R,\theta,\varphi)=-\frac{u}{r_{12}}-\frac{u}{r_{23}}-\frac{u}{r_{31}},
\end{equation} 
where $u$ is the gravitation-like coupling. The interparticle
distances $r_{ij}$ are given in terms of the hyperspherical
coordinates by     
\begin{eqnarray}
r_{12}&=&
{3^{-1/4}}{R}\left[1+\sin{\theta}\sin(\varphi-{\pi}/{6})\right]^{1/2}, 
\nonumber \\
r_{23}&=&
{3^{-1/4}}{R}\left[1+\sin{\theta}\sin(\varphi-5{\pi}/{6})\right]^{1/2}, 
\nonumber \\
r_{31}&=&
{3^{-1/4}}{R}\left[1+\sin{\theta}\sin(\varphi+{\pi}/{2})\right]^{1/2}. 
\end{eqnarray}

\begin{figure}
\includegraphics[width=2.3in,angle=270,clip=true]{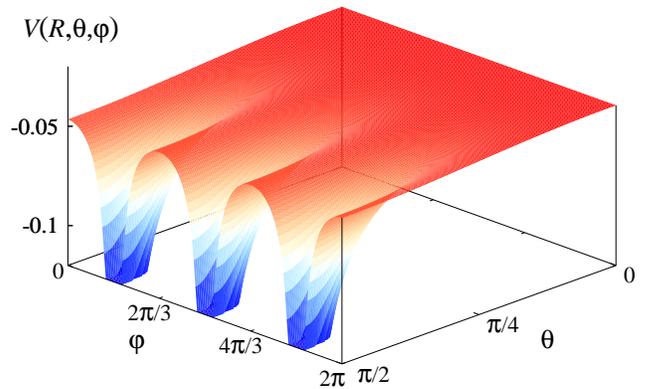}
\caption{The potential $V(R,\theta,\varphi)$ at
  $R=100$. Due to the symmetry properties of the three identical
  particles, the adiabatic equation [Eq.~(\ref{poteq})] is solved only for
  $\theta=0$ to $\pi/2$ and $\varphi=0$ to $2\pi/3$} 
\label{Potential}
\end{figure}

Figure~\ref{Potential} shows the potential $V(R,\theta,\varphi)$ as a
function of $\theta$ and $\varphi$ at 
$R=100$. The singular points at $\theta=\pi/2$ and
$\varphi=\pi/3$, $\pi$ and $5\pi/3$ are the points where $r_{23}=0$,
$r_{31}=0$ and $r_{12}=0$, respectively. Notice that our choice for
the hyperangles $\theta$ and $\varphi$ \cite{Esry-02,Esry-03} --- like any of
the so-caller democratic or Smith-Whitten coordinates --- implies a 
periodicity of the potential in $\varphi$ for identical particles,
which substantially simplifies the numerical solution of
Eq.~(\ref{poteq}) when symmetrizing the wave function. The 
two-dimensional equation for $J^{\pi}=0^{+}$ thus needs to be solved
only from $\varphi=0$ to $2\pi/3$, with the requirement that the
derivative of $\phi(R;\theta,\varphi)$ with respect to $\varphi$ is
zero at each boundary, followed by postsymmetrization to extract the
completely symmetric solutions. We have treated the cusp due the
$1/r$ divergence as $r\rightarrow0$ by including more spline functions
\cite{Boor1978} at $\varphi=\pi/3$. For other $J^{\pi}$ the solutions
can be determined in a similar way \cite{Esry-02,Esry-03}. 

In the atomic case, suggested in Ref.~\cite{DODell2000}, $u$ is given
by (in S.I. units) 
\begin{equation}
u=\frac{11}{4\pi}\frac{Iq^2\alpha_{p}^2}{c\varepsilon_{0}^2}
\end{equation}
\noindent
where $I$ is the laser intensity, $q$ the photon wave number, and
$\alpha_{p}$ the atomic dynamic polarizability. So, under the
conditions discussed in Ref.~\cite{DODell2000}, the strength of the
interaction is controllable via the laser's parameters. For the
present calculations, however, we base our units on $u$, thus
producing a unitless equation. Our length units are $2\hbar^2/mu$, and
our energy units are $mu^2/2\hbar^2$. These yield, in analogy to
atomic units, a two-body energy spectrum $E_{n}=-1/2n^2$.

\section{Results and Discussion}

\begin{figure}
\includegraphics[width=2.8in,angle=270,clip=true]{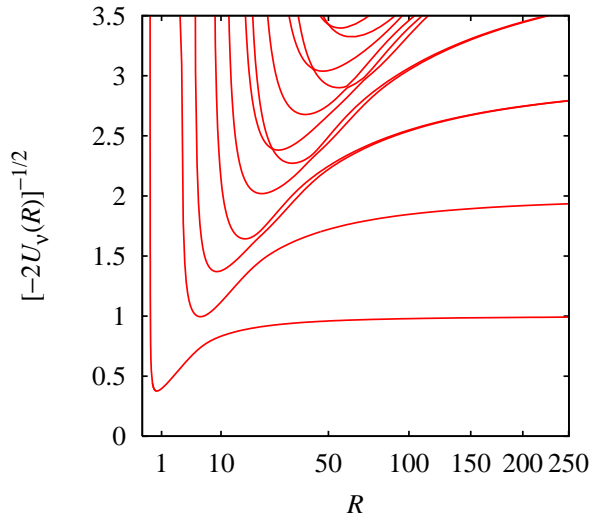}
\caption{The hyperspherical potentials for three identical bosons with
         attractive $1/r$ interactions. For large values of 
         $R$, $[-2U_{\nu}(R)]^{-1/2}$ converges to the principal
         quantum number $n_{\rm 2b}$ for the hydrogen-like subsystems.}  
\label{AdiaPot}
\end{figure}

Figure \ref{AdiaPot} shows the effective three-body potentials in the
form $[-2U_{\nu}(R)]^{-1/2}$ such that for large values of $R$ they
converge to the principal quantum number $n_{\rm 2b}$ associated with the
two-body hydrogen-like subsystems (for $R\rightarrow\infty$ one
particle is far from the others). The lowest potential in
Fig.~\ref{AdiaPot}, converging to $n_{\rm 2b}=1$, supports the $0^+$ bound
states (or $S^e$, in analogy with atomic spectroscopic notation).
In particular, it contains both the ground state and the $S^e$ series
of singly excited states, using the language of atomic structure. The
higher potentials (converging to $n_{\rm 2b}>1$) support series of
doubly excited states that are coupled to and can decay to the
continuum of the lowest potential and are thus metastable. 
These comments draw on the similarity of these potentials to those for
three-body systems like He and H$^-$ \cite{CDLin}. There are minor
differences, of course, due to the different permutational symmetries
of the two systems and to the absence of the Coulomb repulsion for the
present case.  
   
\begin{figure}
\includegraphics[width=2.3in,angle=270,clip=true]{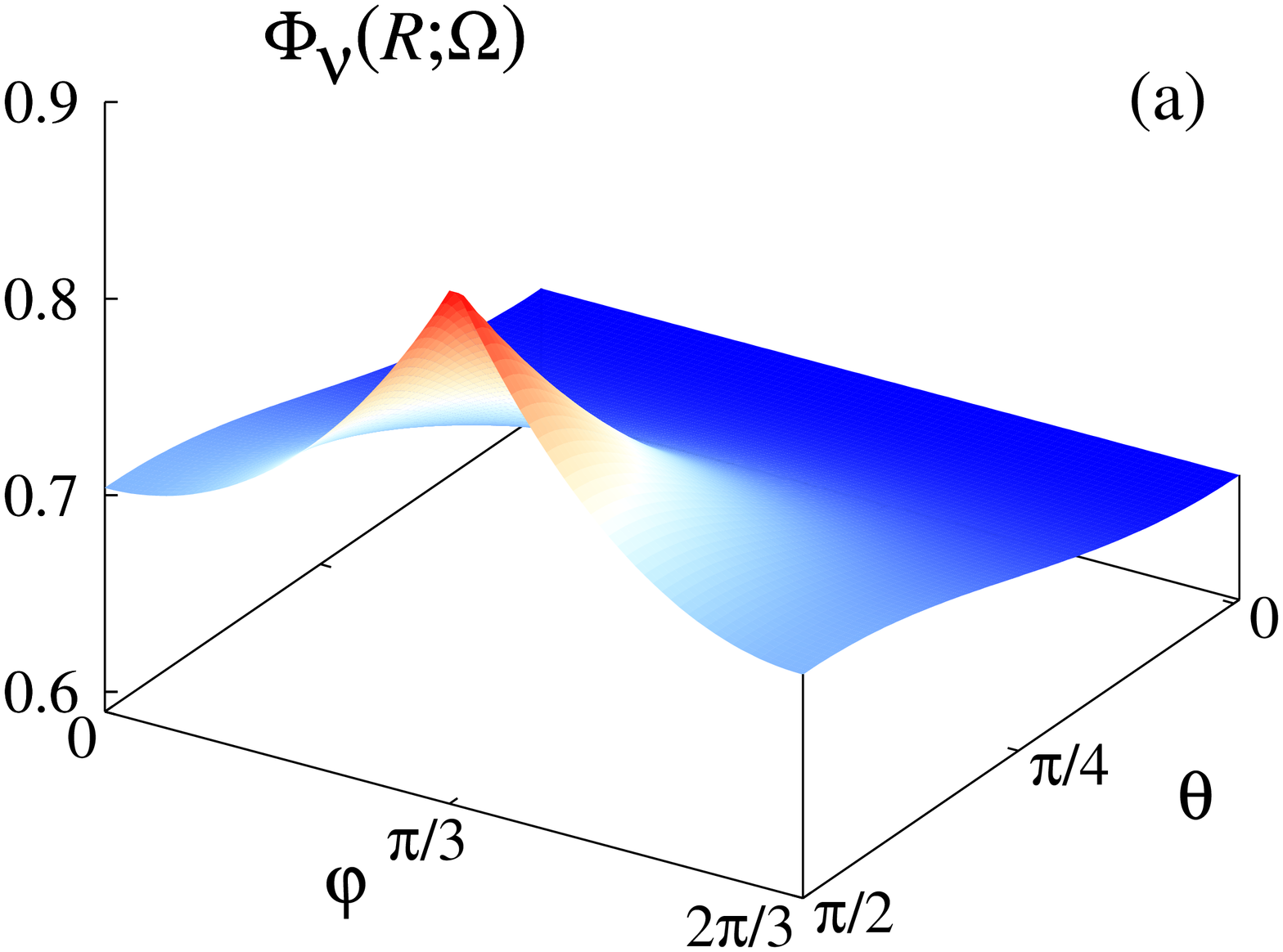}
\includegraphics[width=2.3in,angle=270,clip=true]{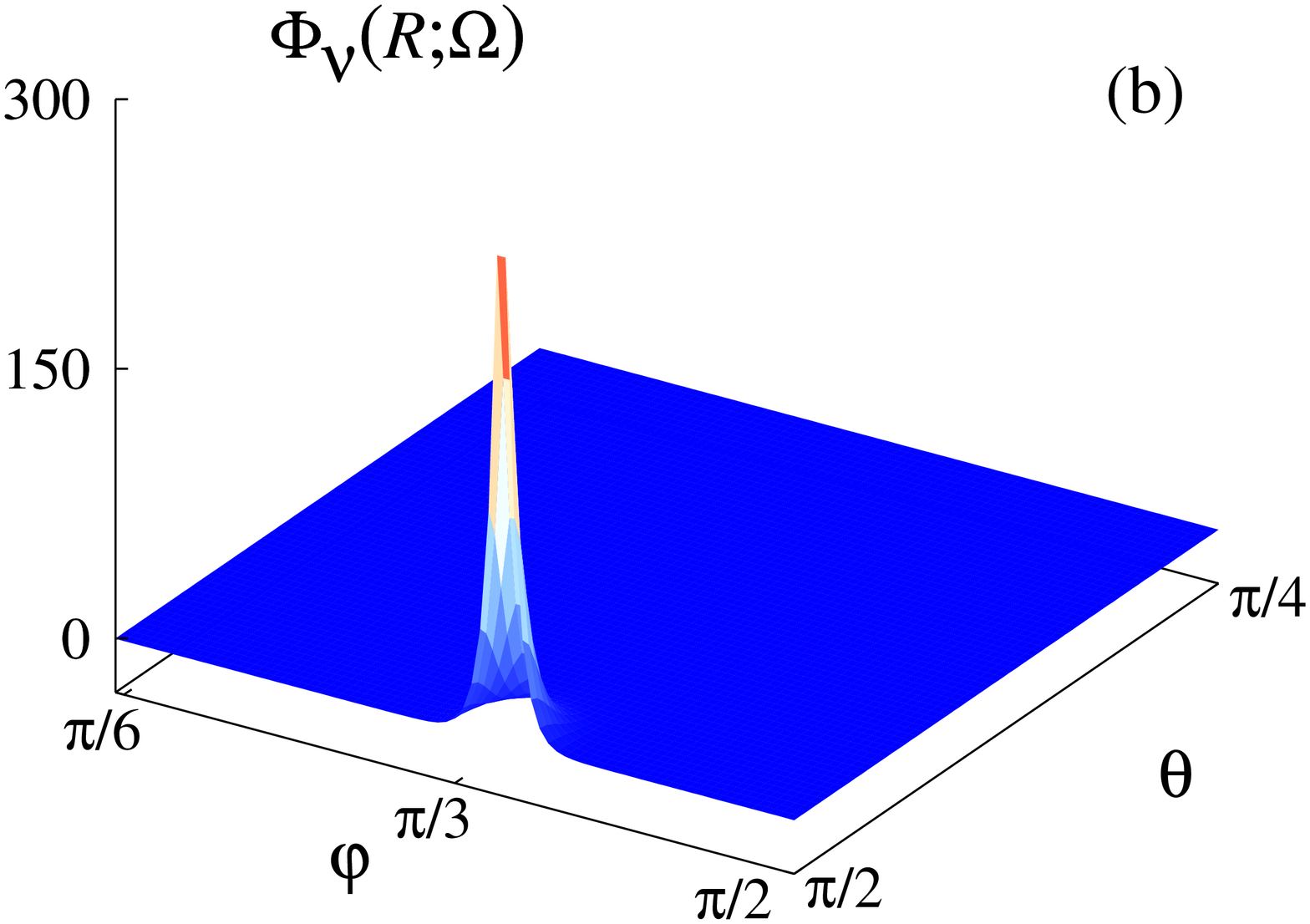}
\caption{The lowest $J^{\pi}=0^+$ channel function ($\nu =1$) as a
  function of $\theta$ and $\varphi$ at (a) $R=0.69$ and (b)
  $R=100$.}
\label{RadWFRminV1}
\end{figure}

\begin{figure}
\includegraphics[width=2.3in,angle=270,clip=true]{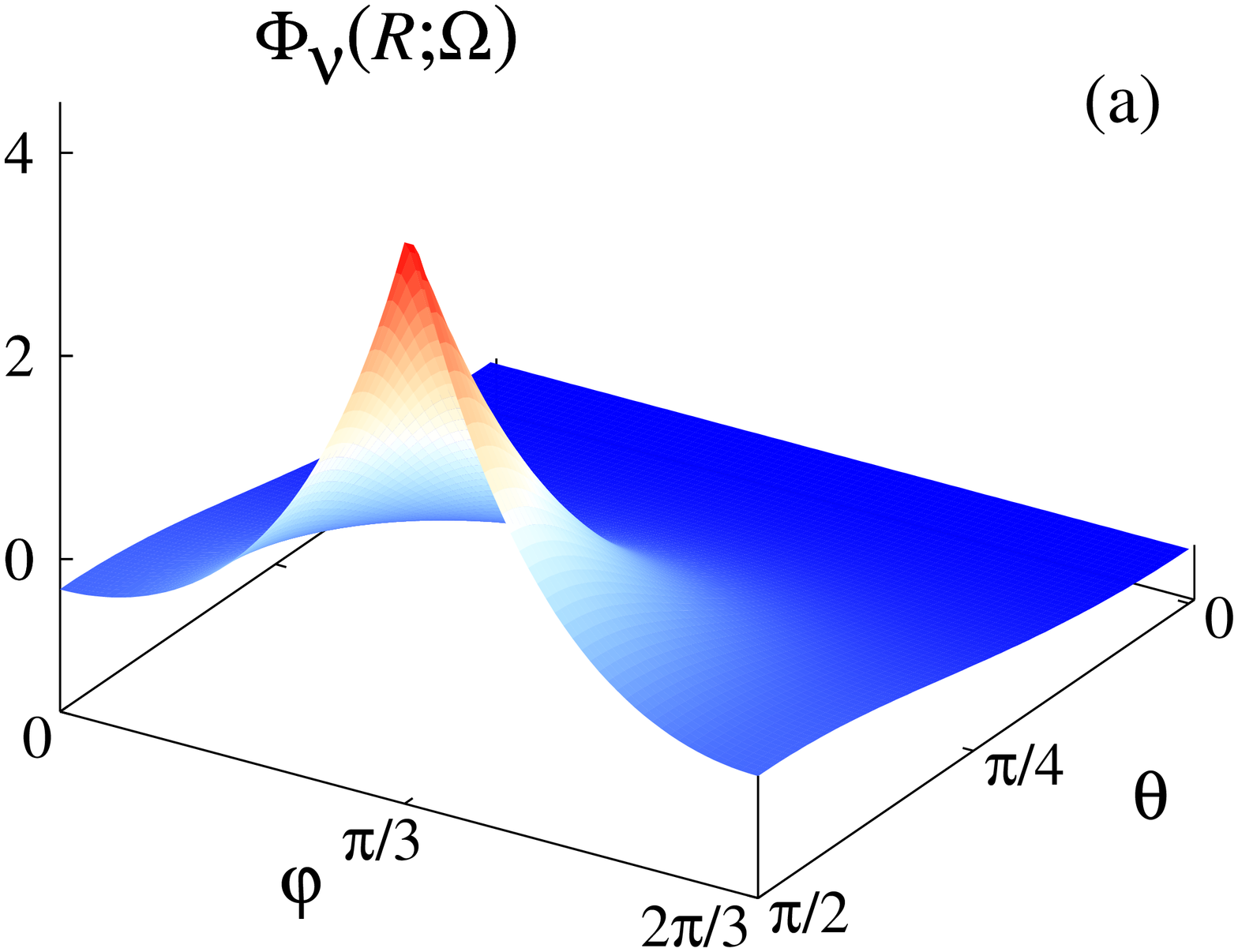}
\includegraphics[width=2.3in,angle=270,clip=true]{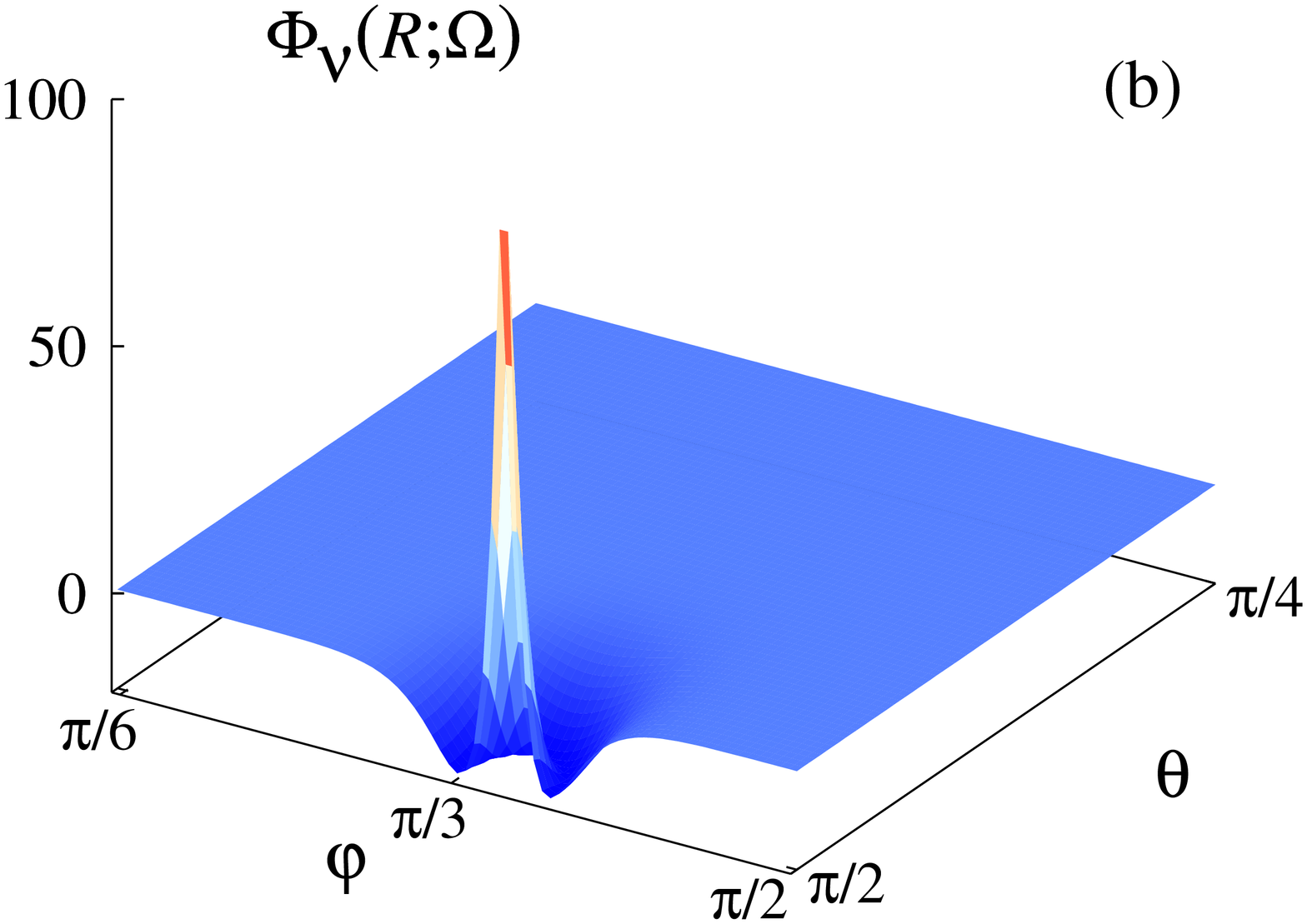}
\caption{The first excited $J^{\pi}=0^+$ channel function ($\nu =2$)
  as a function of $\theta$ and $\varphi$ at (a) $R=5.75$ and (b)
  $R=100$.}
\label{RadWFRminV2}
\end{figure}

Since the $J^{\pi}=0^{+}$ channel functions depend only on the
hyperangles $\theta$ and $\varphi$ they can be plotted in their
entirety for each value of $R$. Figures \ref{RadWFRminV1} and
\ref{RadWFRminV2} show the channel functions for the two lowest
channels at $R=0.69$ and $100$ for $\nu=1$, and $R=5.75$ and
$100$ for $\nu=2$. The first $R$ value lies near the respective
potential minima; and the second, in the asymptotic region.
For small $R$, we plot $\Phi_{\nu}$ in the range $0\le\varphi\le
2\pi/3$ from which the function in the whole range $0\le\varphi\le
2\pi$ can be obtained by symmetry (translation of the plotted portion
by $2\pi/3$ and $4\pi/3$). For large $R$, however, we plotted the
solution only in the range $\pi/6\le\varphi\le\pi/2$ to emphasize the
two-body character of the solution.
 
The hyperangular distributions are useful because they reveal the
geometry of the 
system. At $\theta=0$, for instance, the atoms form an equilateral
triangle, while for $\theta=\pi/2$ they lie along a line. 
Figure \ref{RadWFRminV1}(a) shows that near the $\nu=1$ potential
minimum (see Fig.~\ref{AdiaPot}) the lowest channel function is spread
out over the entire hyperangular plane with an increased
amplitude near the two-body coalescence point ($r_{31}=0$) at
$\theta=\pi/2$ and $\varphi=\pi/3$. This point corresponds to a linear
configuration with two of the particles closer to each other than to
the third particle, representing strong two-body correlations. Since
this $R$ is the equilibrium distance, the three particles in the
ground state thus assume all triangular shapes between linear and
equilateral with a preference for linear shapes having two particles
close to each other (taking into account the $sin2\theta$ volume
element). As $R$ increases [Fig.~\ref{RadWFRminV1}(b)] the 
channel function ``collapses'' to the region around the coalescence
points, displaying the two-body character it must have for
$R\rightarrow\infty$ --- in this case the $1s$ state. Figures
\ref{RadWFRminV2}(a) and \ref{RadWFRminV2}(b) show the $\nu=2$ channel
function. They show much the same behavior as $\nu=1$ except with the
necessary addition of a node. Figure \ref{RadWFRminV2}(b) shows that
$\nu=2$ converges to the $2s$ state of the two-body subsystem. Note
that for identical spinless (or spin-stretched) bosons, $2p$ two-body
states are not allowed by symmetry, so there is only a single
potential correlating to $n_{\rm 2b}=2$ at $R\rightarrow\infty$.

\begin{figure}
\includegraphics[width=2.25in,angle=270,clip=true]{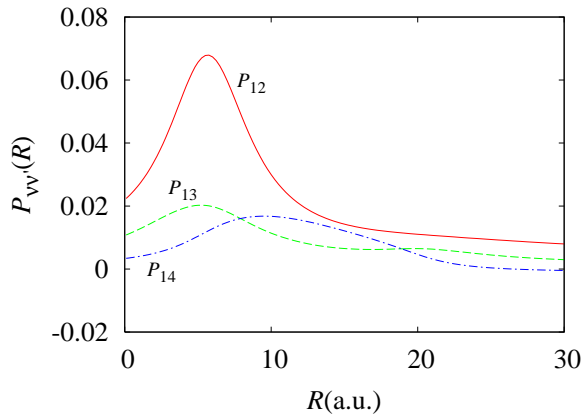}
\caption{Nonadiabatic coupling $P_{\nu \nu \prime}$ between the 
  lowest channel and the next three channels.}
\label{Coupling}
\end{figure}

In Figure~\ref{Coupling} we show the nonadiabatic couplings 
$P_{\nu\nu'}(R)$ between the lowest channel and the next three
channels ($P_{\nu\nu}=0$ and $P_{\nu\nu'}=-P_{\nu'\nu}$) calculated
from Eq.~(\ref{quv}). The fact that the coupling $P_{12}$ in
Fig.~\ref{Coupling} is substantially larger than the couplings with
higher channels implies rapid convergence for the bound state energies
as a function of the number of channels. Further, $P_{12}$ peaks
around $R=6$ which correlates roughly with the location of an avoided
crossing between the corresponding potential curves as expected (see
Fig.\ref{AdiaPot}). 

\begin{figure}
\includegraphics[width=2.25in,angle=270,clip=true]{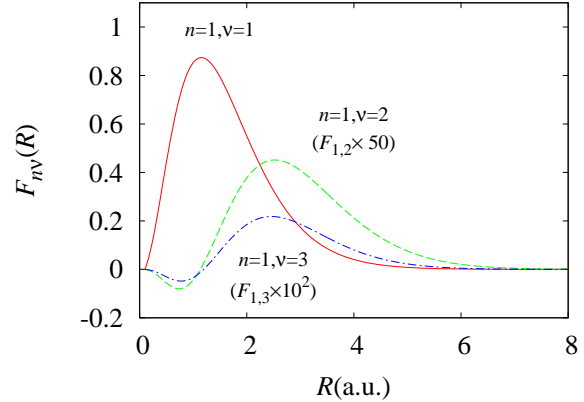}
\caption{First three components of the ground state hyperradial
   wavefunction $F_{n\nu}$ ($n=1$) associated with the first three
   channels $\nu=1$, $2$, and $3$.}  
\label{WF}
\end{figure}

We have solved Eq.~(\ref{radeq}) and determined the bound state
energies and hyperradial wavefunctions $F_{n\nu}(R)$ including up to
15 channels. For example, Figure~\ref{WF} shows the ground state wave
function for a three channel calculation.
The $\nu=1$ component is associated with the lowest adiabatic channel,
while the $\nu=2$ and $\nu=3$ are related to the next two adiabatic   
channels and are present due to the coupling between the
channels. Note that the probability given by 
${\cal P}_{n\nu}=\int_{0}^{\infty}|F_{n\nu}(R)|^2dR$ due to the first
term dominates the other contributions. In Table~\ref{ProRadialWF} we
show ${\cal P}_{n\nu=1}$ for the ground state ($n=1$) and
the four lowest excited states for a 15 channel calculation. The
$\nu=1$ adiabatic channel represents roughly $99\%$ or more of the 
probability for each state, showing that the adiabatic expansion is,
in fact, quite good.
\begin{table}[htbp]
\begin{ruledtabular}
\caption{The probability ${\cal P}_{n\nu}$ associated with the $\nu=1$
  term of the expansion (\ref{chfun}) for the ground state ($n=1$) and
  the next four excited states.\label{ProRadialWF}}  
\begin{tabular}{cccc}
  $n$  &  ${\cal P}_{n1}=\int_{0}^{\infty}|F_{n1}|^2dR$   \\ \hline
   1        &  99.98$\%$   \\
   2        &  99.80$\%$   \\
   3        &  99.15$\%$   \\
   4        &  98.90$\%$   \\
   5        &  99.21$\%$   \\
\end{tabular}
\end{ruledtabular}
\end{table}
 
In Table~\ref{EnergyChannels} we show the ground state energy as a
function of the number of channels, further demonstrating the expected
rapid convergence of the adiabatic expansion. The ground state energy
for this system has certanly been calculated before. In
Ref.~\cite{JLBasdevant}, for instance, a ground 
state energy of $E_0\cong -1.067\:G^2m^5/\hbar^2=-2.134\:mu^2/\hbar^2$
(in our units) was obtained. Comparison with
Table~\ref{EnergyChannels} shows that since both calculations are
variational, our single channel calculation already gives a more
precise result. We speculate that the large differences in the
potential energies shown in Fig.~\ref{AdiaPot} are the main reason
that a single channel already gives such a good result. In fact, this
channel separation is closely related to  
the small magnitude of the coupling terms shown in Fig.~\ref{Coupling}. 
Table~\ref{EnergyChannels} also shows that our six channel
approximation for the ground state gives a result converged to seven   
digits.
\begin{table}[htbp]
\caption{Convergence of the ground state energy as a function of the 
number of channels included in Eq.~(\ref{radeq}).\label{EnergyChannels}} 
\begin{ruledtabular}
\begin{tabular}{cccc}
  Number of channels  &    Ground state energy ($\:mu^2/\hbar^2$) &  \\ \hline
  1      & -2.136\:033   &  \\
  2      & -2.136\:481   &  \\
  3      & -2.136\:523   &  \\
  4      & -2.136\:525   &  \\
  5      & -2.136\:526   &  \\
  6      & -2.136\:527   &  \\
  15     & -2.136\:527   &  \\
\end{tabular}
\end{ruledtabular}
\end{table}

In Table~\ref{EnergyBoundn} we give our converged results for the
ground and first four excited states (calculated with 15 channels). 
Is well known \cite{Starace} that the hyperspherical energy obtained 
disregarding all couplings and the energy obtained considering only
the diagonal coupling in Eq.~(\ref{radeq}) are, in fact, lower and upper
bounds for the ground state energy, respectively. In these
approximations, we 
obtained a lower bound corresponding to $-2.138\:650\:mu^2/\hbar^2$
and an upper bound of $-2.136\:033\:mu^2/\hbar^2$. 
The difference between them is about $0.1\%$
while the results obtained in Ref.~\cite{JLBasdevant} give a
difference of about $10\%$. By comparison, for a system like the He
atom \cite{Groote}, where the electronic repulsion plays an important
rule, the relative difference between the lower and upper bounds
estimated from hyperspherical potential curves is about $1\%$.

\begin{table}[htbp]
\caption{Ground state and excited states energies $E_{n,\nu}$
  ($\nu=1$) calculated using 15 coupled channels.\label{EnergyBoundn}} 
\begin{ruledtabular}
\begin{tabular}{cccc}
  $n$     &    $E_{n,\nu}$ ($\:mu^2/\hbar^2$)  \\ \hline
   1      & -2.136\:527    \\
   2      & -1.145\:881    \\
   3      & -0.786\:454    \\
   4      & -0.661\:162    \\
   5      & -0.603\:740    \\
\end{tabular}
\end{ruledtabular}
\end{table}

\section{summary}

We have used the adiabatic hyperspherical representation to describe
system of three identical bosons with attractive $1/r$
potentials. Such a system might eventually be created experimentally
by irradiating ultracold atoms with intense, extremely off-resonant
lasers. 
We calculated the ground state and excited state energies converged
to seven digits which represents a substantial improvement over
previous results. Our method is essentially exact, with the  
only approximation being the truncation of the number of channels
used in the expansion of the total wave function. 
Although other methods, such as Hylleraas
variational techniques, might provide much better
bound states energies, as we have shown here, the adiabatic
hyperspherical representation naturally offers qualitative information
along with quantitative results.

\acknowledgments
This work was supported by the National Science Foundation.

\end{document}